\documentclass[doublecol]{epl2}

\title{Coherent laser control of the current through molecular junctions} 
\shorttitle{Coherent laser control of the current}
\author{Guangqi Li \inst{1} \and  Michael
  Schreiber  \inst{2} \and Ulrich Kleinekath\"ofer \inst{1}
 }
\shortauthor{GuangQi Li \etal}
 \institute{
  \inst{1}
School of Engineering and Science, Jacobs University Bremen
\thanks{Formerly International University Bremen},  Campus Ring
1, 28759 Bremen, Germany\\
  \inst{2}
 Institut f\"ur Physik, Technische Universit\"at Chemnitz, 09107
  Chemnitz, Germany  
}

\abstract{
  The electron tunneling through a molecular junction modeled by a single
  site weakly coupled to two leads is studied in the presence of a
  time-dependent external field using a master equation approach. In the
  case of small bias voltages and high carrier frequencies of the external
  field, we observe the phenomenon of coherent destruction of tunneling,
  i.e.\ the current through the molecular junction vanishes completely for
  certain parameters of the external field. In previous studies the
  tunneling within isolated and open multi-site systems was suppressed; it
  is shown here that the tunneling between a single site and electronic
  reservoirs, i.e.\ the leads, can be suppressed as well.  For larger bias
  voltages the current does not vanish any more since further tunneling
  channels participate in the electron conduction and we also observe
  photon-assisted tunneling which leads to steps in the current-voltage
  characteristics.  The described phenomena are demonstrated not only for
  monochromatic fields but also for laser pulses and therefore could be
  used for ultrafast optical switching of the current through molecular
  junctions.
}

\begin{document}

\maketitle

%\baselineskip=23pt

%%%%%=============================================================
%%%%
\section{Introduction}
Electronic transport through molecular wires and junctions has recently
attracted much attention experimentally as well as theoretically
\cite{hang02b,nitz03a,joac04}. Under the influence of a bias voltage and
because of the coupling to the leads which act as electron source and
drain, a current through the molecular junction is established.  When an
external time-dependent field, such as a laser field or an additional ac
voltage is applied to the molecular junction, several interesting effects
arise. One phenomenon is the well-known photon-assisted tunneling (PAT)
\cite{plat04a}. It was studied already in the early 1960's experimentally
by Dayem and Martin \cite{daye62} and theoretically by Tien and Gordon
using a simple theory which captures already the main physics of
PAT\cite{tien63}. The main idea is that an external field periodic in time
with frequency $\omega$ can induce inelastic tunneling events when the
electrons exchange energy quanta $\hbar \omega$ with the external field.
Another important effect is the famous phenomenon of coherent destruction
of tunneling (CDT). Grossmann et al.\ \cite{gros91,gros91b,gros92} first
studied this effect and found that tunneling can be quenched in a
periodically driven quantum system.  In the context of molecular wires,
this phenomenon can be explained using Floquet theory in the case of a
periodic laser field \cite{lehm03a}, and CDT occurs for certain amplitudes
of the laser field at fixed frequencies
\cite{cama03,cama04,kohl04a,kohl04b,wela05a}. Different scenarios of
controlling the tunneling in molecular wires and quantum dots have been
proposed based on different mechanisms
\cite{bran00,cref04,wela05a,kohl04a,plat04a,cota05}. Also current-induced
light emission in molecular junctions has been studied \cite{galp05}.

In the current paper we focus on the tunneling through a single-site
molecular junction. This might be a quantum dot (though the temperatures in
the current examples are rather high for quantum dots) or a single
molecular level acting as a molecular wire. The theoretical foundation is a
density matrix formalism using a perturbative treatment within the
molecule-lead coupling to second order. Applying this technique it is
possible to calculate the time-dependent population in and the current
through the molecular junction under the influence of a time-varying
external field \cite{wela05a,klei06b}. Since the effect of the external
field on the coupling between molecule and leads is treated exactly and not
neglected as for example in Redfield theory, effects based on the influence
of the laser on this coupling can be investigated.  The tunneling between
the molecule and the leads can, for small bias voltages, be suppressed by a
monochromatic laser. This result is different from previous studies
\cite{kohl04b,kohl04a,wela05a,klei06c} in which the current vanished
because of the CDT between the sites of the wire. In Ref.~\cite{kohl04b}
the possibility of CDT for a single site was briefly mentioned as a
limiting case of a wire with two sites having a large intersite coupling,
but it was not further explored.  Additionally we demonstrate that not only
monochromatic laser fields but also laser pulses can lead to CDT. This
opens opportunities for building optical current switches with a time
resolution on the femtosecond scale. $\hbar$=1 is used throughout the rest
of this Letter.

\section{Model}
The total system Hamiltonian $H(t)=H_S(t)+H_R+H_{SR}$ includes three
parts: the relevant system $H_S(t)$  mimicking the molecule, the reservoirs
$H_R$ modeling the two leads and the system-reservoir coupling $H_{SR}$.
The creation and annihilation operators of electrons with spin $\sigma$ are
denoted by $c^\dagger_{\sigma}$ and $c_{\sigma}$, respectively, so that the
description of a single-site molecule reads
\begin{eqnarray} \label{equ:Ham_wire}
H_S(t)&=&\sum_{\sigma} \left(\varepsilon_0 - \mu{}E(t) \right)
c_{\sigma}^\dagger c_{\sigma} 
%- \Delta\sum_{n\sigma} ( c_{n\sigma}^\dagger c_{n+1,\sigma}+c_{n+1,\sigma}^\dagger c_{n\sigma}) \nonumber
+ U c_{\uparrow}^\dagger c_{\uparrow}  
c_{\downarrow}^\dagger c_{\downarrow} 
%\sum_{\sigma} c_{\sigma}^\dagger c_{\sigma}  
\end{eqnarray}
with spin-independent on-site energy $\varepsilon_0$ and electron
interaction $U$ within the doubly occupied states. The time-dependent term
$-\mu{}E(t)$ describes the effect of the external field $E(t)$ and for
simplicity we assume that the proportionality factor $-\mu{}$ equals unity.
The external field is of the form $E(t)=E_0(t) \cos\omega t$ with a
possibly time-varying amplitude $E_0(t)$. The two leads coupled to the
molecular junction are mimicked as electron reservoirs in thermal
equilibrium by $H_R=\sum_{q\sigma} \omega_{q\sigma} b_{q\sigma}^\dagger
b_{q\sigma} $.  Here $b_{q\sigma}^\dagger$ and $b_{q\sigma}$ denote the
creation and annihilation operators of an electron with spin $\sigma$ in
reservoir mode $\omega_q$. Due to the assumed thermal equilibrium of the
electronic leads, the occupation expectation values of the reservoir modes
are determined by $ \label{equ:equi} \langle b_{q\sigma}^\dagger
b_{q'\sigma} \rangle = n_F(\omega_q-E_F) \delta_{qq'}$ where $n_F$ is the
Fermi function and $E_F$ the Fermi energy. In further derivations we will
only refer to the left lead but the formalism has to be applied to the
right lead coupled to the wire as well.  The coupling of the left lead to
the molecule is given by
\begin{equation} \label{equ:Ham_coup}
H_{SR}= \sum_{\sigma,x=1,2} K_{x\sigma}  \Phi_{x\sigma} 
=  \sum_{q\sigma} (V_q c_{1\sigma}^\dagger b_{q\sigma} 
+ V_q^* b_{q\sigma}^\dagger c_{1\sigma})
\end{equation}
with $\Phi_{1\sigma}=\Phi_{2\sigma}^{\dagger}=\sum_q V_q b_{q\sigma}$,
$K_{1\sigma}=K_{2\sigma}^{\dagger}=c_{1\sigma}^\dagger$, and a wire-lead
coupling strength $V_q$. For the coupling of the molecule to the right lead
similar equations hold.  

The calculations are performed with and without electron interaction in the
molecule.  Neglecting electron interaction all electron-carrying states of
the molecule are degenerate with energy $\varepsilon_0$. In the case of
electron interaction $U$, double occupancy leads to a state with energy $2
\varepsilon_0+U$.  When the bias voltage is small and $\varepsilon_0$ is
between the Fermi energies $E_{F,r}$ and $E_{F,l}$ of the right and the
left lead, respectively, only the tunneling through the channel with energy
$\varepsilon_0$ leads a current, $I_0$.  When the value of the bias voltage
is above $U$, an extra channel opens and there is a shoulder in the
current.  Thielmann et al.  \cite{thie03} used these energy levels to
explain the shoulders in the tunneling current for a quantum dot at low
temperature.  The excited state in the investigations by Bruder
\cite{brud94,staf96}, Oosterkamp \cite{oost96,oost97,oost98} and others has
the same effect\cite{sun98,qin01}.

Since in most cases one is only interested in the time evolution of the
relevant system, i.e.\ in this case the molecule, a time-local quantum master
equation (QME) based on a second-order perturbation theory in the
molecule-lead coupling has been developed for the reduced density matrix of
the molecule $\rho_S(t)$ \cite{wela05a,klei06b}
\begin{eqnarray}\label{equ:master2local}
\frac{\partial\rho_S(t)}{\partial t}&=&-i \mathcal L_S(t) \rho_S(t) \\
&& -\sum_{\sigma xx'} [K_{x\sigma},\, \Lambda_{xx'\sigma}(t)\rho_S(t)-
\rho_S(t)\widehat \Lambda_{xx'\sigma}(t)] \nonumber
\end{eqnarray}
with auxiliary operators for the molecule-lead coupling
\begin{eqnarray}\label{equ:aux1local}
\Lambda_{xx'\sigma}^{}(t) = \int_{t_0}^t \mathrm dt' C_{xx'}(t-t')  
U_S(t,t')  K_{x'\sigma},\\ %\nonumber \\
\widehat \Lambda_{xx'\sigma}(t) = \int_{t_0}^t \mathrm dt' C_{x'x}^*(t-t') 
 U_S(t,t')  K_{x'\sigma}~.
\end{eqnarray}
Here $\mathcal L_S(\tau)= [H_S(\tau),\bullet]$ is the Liouville operator,
$U_S(t,t')=T_+\exp\{-i \int_{t'}^t \mathrm d\tau \mathcal L_S(\tau)\}$ the
time evolution operator and $C_{x x'}(t)=\mathrm{tr}_R \lbrace {\rm e}^{i
  H_R t} \Phi_x {\rm e}^{-i H_R t} \Phi_{x'} \rho_R \rbrace$ the reservoir
correlation functions with spin-independent reservoir density matrix
$\rho_R$. $T_+$ is the time-ordering operator. 

The properties of the Fermionic reservoirs are described by a single
quantity, the spectral density $J_R(\omega)$.  For a dense spectrum,
$J_R(\omega)$ is a smooth function and one can approximate it by a
numerical decomposition into few Lorentzian functions \cite{wela05a}
\begin{equation} \label{equ:spectralnum}
J_{R}(\omega)=\sum_{k=1}^m \frac{p_k}{4 \Omega_k}  
\frac{1}{(\omega -\Omega_k)^2+\Gamma_k^2}
\end{equation}
with real fitting parameters $p_k$, $\Omega_k$ and $\Gamma_k$. 
Using the theorem of residues and denoting the Fermi function as $n_F$
yields
\begin{eqnarray} \label{bath12dec}
C_{12}(t)&=&\sum_{k=1}^m \frac{p_k}{4 \Omega_k \Gamma_k}
\left(n_F(-\Omega_k^- +E_F) {\rm e}^{-i\Omega_k^- t} \right) \nonumber \\
&&-\frac{2i}{\beta} \sum_{k=1}^{m'} J_{R}(\nu_k^\ast) {\rm e}^{-i \nu_k^\ast t}
=\sum_{k=1}^{m+m'} a_{12}^k {\rm e}^{\gamma_{12}^k t}   \\ 
C_{21}(t)&=& \sum_{k=1}^m \frac{p_k}{4 \Omega_k \Gamma_k}
\left(n_F(\Omega_k^+-E_F)  {\rm e}^{i\Omega_k^+ t} \right) \nonumber \\
&&-\frac{2i}{\beta} \sum_{k=1}^{m'} J_{R}(\nu_k) {\rm e}^{i \nu_k t}
=\sum_{k=1}^{m+m'} a_{21}^k {\rm e}^{\gamma_{21}^k t}
\end{eqnarray}
with the abbreviation $\Omega_k^\pm{}=\Omega_k \pm{}i \Gamma_k$ and the Matsubara
frequencies $\nu_k=i\frac{2\pi k + \pi}{\beta} +E_F$.  The infinite sums
over the $\nu_k$  can be truncated at a finite value depending
on the temperature $T$ and the spectral width of $J_R(\omega)$. With these
forms of $C_{12}$ and $C_{21}$ one can obtain a set of differential
equations for the auxiliary density operators
\begin{eqnarray}\label{equ:diffaux}
\frac{\partial}{\partial t} \Lambda_{xx'\sigma}^k(t)&=& a_{xx'}^k  K_{x'\sigma} 
-i [H_S(t), \Lambda_{xx'\sigma}^k(t)] \nonumber \\
&&+ \gamma_{xx'}^k \Lambda_{xx'\sigma}^k(t), \\
\frac{\partial}{\partial
  t}{\widehat\Lambda}_{xx'\sigma}^k(t)&=&\left(a_{x'x}^k\right)^\ast
K_{x'\sigma}   -i [H_S(t), \widehat \Lambda_{xx'\sigma}^k(t)] \nonumber \\ 
\label{equ:diffaux2}
&&+ \left(\gamma_{x'x}^k \right)^\ast \widehat \Lambda_{xx'\sigma}^k(t)
\end{eqnarray}
with ${\Lambda}_{xx'\sigma}(t)=\sum_{k=1}^{m+m'}
{\Lambda}_{xx'\sigma}^k(t)$ and
${\widehat\Lambda}_{xx'\sigma}(t)=\sum_{k=1}^{m+m'}
{\widehat\Lambda}_{xx'\sigma}^k(t)$.  As has been  detailed previously
\cite{wela05a}, the coupled differential Eqs.~(\ref{equ:master2local},
\ref{equ:diffaux}, \ref{equ:diffaux2}) can now be numerically integrated
applying, e.g., the Runge-Kutta scheme.

Using the electron number operator of
the left lead with the summation performed over the reservoir degrees of
freedom $N_l=\sum_{q\sigma} b_{q\sigma}^{\dagger} b_{q\sigma}$, the
expression for the current is given by \cite{wela05a}
\begin{eqnarray}\label{equ:finalcurrent}
I_l(t)&=&e\frac{\mathrm d}{\mathrm dt} \mathrm{tr} \, \lbrace N_l \rho_S(t)
\rbrace =-ie \, \mathrm{tr} \left\{ [N_l,H(t)] \rho_S(t) \right\} \nonumber \\
&&\hspace{-1.7cm} =2e \, \mathrm{Re} \left(\mathrm{tr}_{S} \left\{ c_{1\sigma}^\dagger \Lambda_{12\sigma}(t)\rho_S(t) -c_{1\sigma}^\dagger \rho_S(t) \widehat\Lambda_{12\sigma}(t) \right\} \right)~.
\end{eqnarray}
Here $e$ denotes the elementary charge. This equation describes the current
$I_l(t)$ from the left lead into the molecule. A similar expression holds
for $I_r(t)$ from the right lead into the molecule.  In a steady state and
after averaging over one period of the driving field, $I_l$ and $I_r$ have
the same magnitude but opposite signs and therefore a total transient
current through the molecular junction can be defined as
$I(t)=(I_l(t)-I_r(t))/2$.  The time-dependent average current $\bar{I}$
will be determined below by averaging $I(t)$ over five periods of the
highly oscillating carrier field.

A simple spectral density with only one Lorentzian ($m$=1) was chosen. With
$\Omega_1=\varepsilon_0$ we locate the maximum of the coupling at the site
energy. By using $\Gamma_1=5 \omega$, the coupling between the leads and
the system is almost in the wide-band limit.  Choosing $p_1=0.04 \omega
\Omega_1 \Gamma_1^2$ we obtain a maximum coupling strength of 0.01 $\omega$
which is smaller than the thermal energy $k_B T$ = $0.025\omega$ and much
smaller than the external field energy $\omega$. So the results below are
all within the high-frequency limit.
%Furthermore one uses the fact that $J_R(\omega)\approx 0$ for $\omega \le 0$ which is
%fulfilled for a wide range of parameters $\Omega_k$ and $\Gamma_k$.
%With the  roots of $n_F$ and (\ref{equ:spectralnum}), 
%the application of the theorem of residues results in

According to the approximate Tien-Gordon model \cite{tien63,plat04a} for
monochromatic external fields which is based on a simplified scattering
picture, the rectified dc currents through ac-driven molecular junctions
are determined as \cite{plat04a}
\begin{equation}\label{equ:tgcurrent}
I_{TG}=\sum_{n=-\infty}^{\infty}J_n^2\left(\frac{E_0}{\omega}\right)
I_{dc}^0(\varepsilon_0+n\omega)= \sum_{n=0}^{\infty} I_n
\end{equation}
where the current in the driven system is expressed by a sum over
contributions of the current $I_{dc}^0(\varepsilon_0+n\omega)$ in the un-driven
case but evaluated at side-band energies $\varepsilon_0+n\omega$ shifted by
integer multiples of the photon quantum and weighted with squares of Bessel
functions. Note that the partial currents $I_n$ contain contributions from
$\pm{}n$.  The term $J_n(E_0/\omega)$ denotes the $n$th-order Bessel function
of the first kind.  The photon absorption ($n>0$) and emission ($n<0$)
processes can be viewed as creating effective electron densities at
energies $\varepsilon_0\pm{}n\omega$ with probability $J_n^2(E_0/\omega)$.
Already Dayem and Martin \cite{daye62} showed clear steps in the tunneling
current induced by these two processes.  Although "a great deal of
information can be extracted from simple models like the Tien-Gordon model"
\cite{plat04a} in many cases one needs more sophisticated approaches.  In
double quantum dot theories, for example, the Tien-Gordon formula has been
recovered in the limit of weak inter-dot coupling
\cite{plat04a,bran04,bran05,bran06}.

In the following we use a bias symmetric with respect to $\varepsilon_0$,
i.e.\ $E_{F,l}=\varepsilon_0+V_b/2$ and $E_{F,r}=\varepsilon_0-V_b/2$.  The
current $I_{TG}$ obtained with the help of the Tien-Gordon theory is
compared to the average current $\bar{I}(t)$ obtained from the QME
described above.  The partial currents $I_n$ defined in
Eq.~(\ref{equ:tgcurrent}) are used to qualitatively explain the steps in
the $I$-$V$ characteristics.  In the results shown below we first restrict
ourselves to a monochromatic laser field with constant field amplitude of
2.405 $\omega$, i.e.\ a zero of the zeroth-order Bessel function
$J_0(E_0/\omega=2.405)$=0.

\begin{figure}
\centerline{
\includegraphics[width=8.5cm,clip]{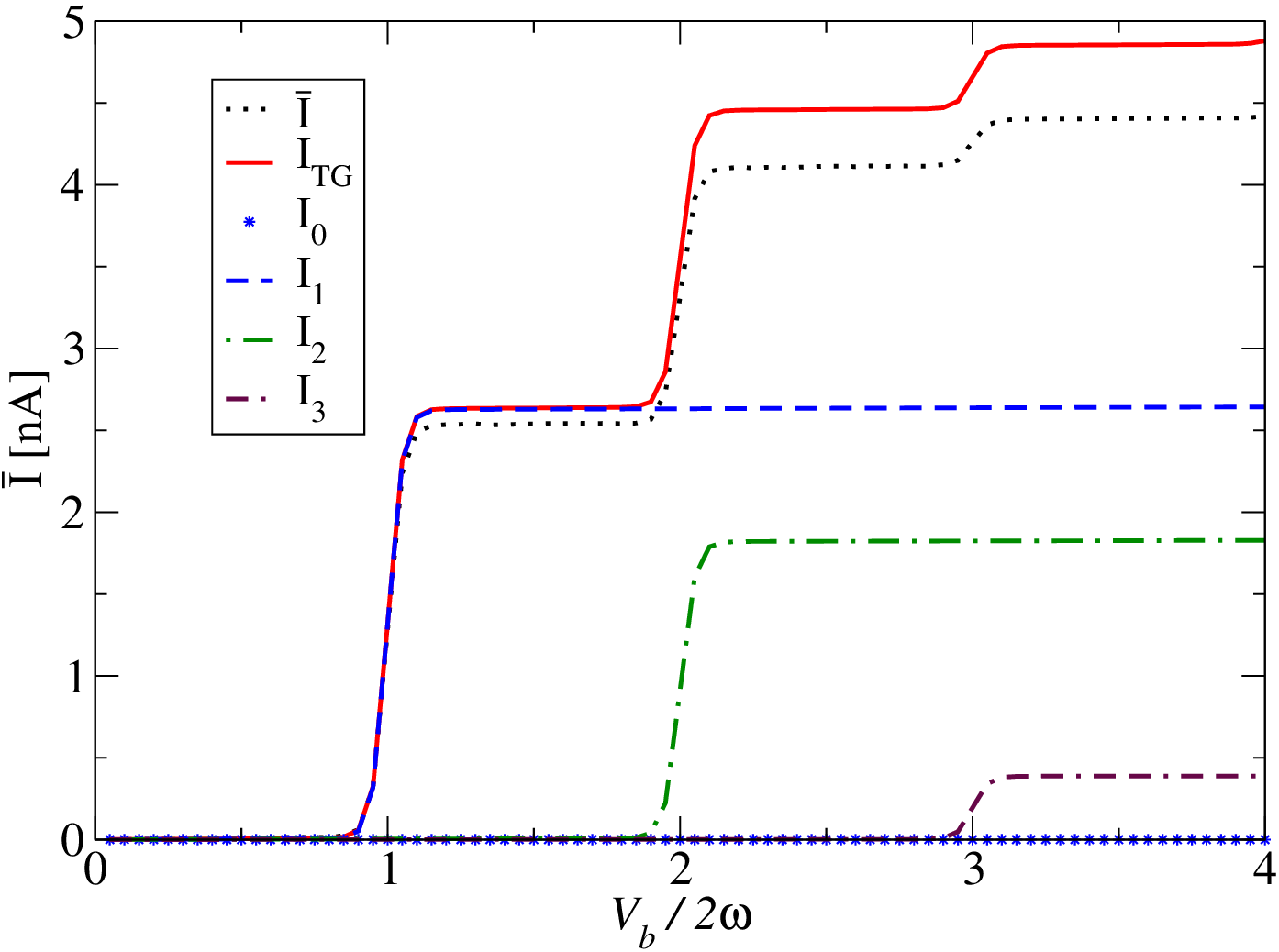}}
\caption{Current induced by a monochromatic laser with amplitude
  $E_0=2.405\omega$.  Average current $\bar{I}$, Tien-Gordon current
  $I_{TG}$, and partial currents $I_0$ to $I_3$ vs $V_b/2\omega$ for $k_B
  T$=0.025$\omega$, $U$=0. From top to bottom: $I_{TG}$, $\bar{I}$, $I_1$,
  $I_2$, $I_3$, $I_0$. Shoulders in the currents are at $V_b/ (2 \omega)$=1, 2,
  3, etc.  }
\label{f.1}
\end{figure}

%\vspace{3ex} 
%\section{ Case $U=0$} 
\section{Monochromatic laser,  $\bm{ U=0}$} 
In this case, energy levels are equal to
$\varepsilon_0$ regardless whether there is already an electron with
opposite spin on the site or not, and lead to the partial current $I_0$.
The PAT-induced states have also equal energies $\varepsilon_0\pm{}n\omega$
leading to the inelastic current contributions $I_n$. In Fig.~\ref{f.1} the
average current $\bar{I}$ shows a step at $V_b/2=\omega$.  When $V_b/2$ is
smaller than $\omega$, the current vanishes.  At these low bias voltages
channels with $n \ne 0$ cannot participate in the conduction since the
energies of the PAT-induced states are not in the window between the Fermi
energies of the right and left contacts.  Because $E_0/\omega=2.405$, $I_0$
also vanishes due to CDT. For other ratios of $E_0/\omega$ it is non-zero.
When $V_b/2=\omega$, the left Fermi energy $E_{F,l} = \varepsilon_0+\omega$
equals the PAT-induced energy and $E_{F,r}=\varepsilon_0-\omega$.
Therefore the current $I_1$ jumps to a finite value, so do $I_{TG}$ and
$\bar{I}$.  A comparison of $I_{TG}$ and $\bar{I}$ in Fig.~\ref{f.1} shows
that each step in these $I$-$V$ curves is related to an inelastic current
$I_n$.  The slope of the steps is of course directly connected to the
temperature.
%Additionally shown in Fig.~1 are the partial currents for $n$=0,$\dots$, 3.
The sum $I_{TG}$ of all  partial currents for $n$=0,$\dots$, 3
agrees reasonably with the current $\bar{I}$ from the
QME, especially for small bias voltages. 

For the current through two coupled quantum dots the Tien-Gordon result is
perturbative in the interdot coupling, as mentioned above
\cite{bran04,bran05,bran06}. For that case, as in our study, the
Tien-Gordon approximation overestimates the current $I_n$. The deviations
between the average current $\bar{I}$ and the Tien-Gordon current $I_{TG}$
become larger with increasing bias. In the present study and for a small
bias voltage the average current stems only from the main contribution
$I_0$ and for this case the Tien-Gordon results match the present results
for different laser amplitudes $E_0$ (not shown).  For larger bias
voltages, the energetic positions of the steps are equal though the
Tien-Gordon approach seems to overestimate the contributions from the
PAT-induced states $\varepsilon_0\pm{}n\omega$.

%\begin{figure}
%\centerline{
%\includegraphics[width=7.5cm,clip]{24.eps}}
%\caption{Population dynamics for a three-site system without leads. The
%  external field periodically alters the energy of the middle site. For the
%  solid curve $E_0/\omega=2.405$ and $E_0/\omega=1$ for the dashed one.}
%\label{f.2}
%\end{figure}

\begin{figure}
\centerline{
\includegraphics[width=8.5cm,clip]{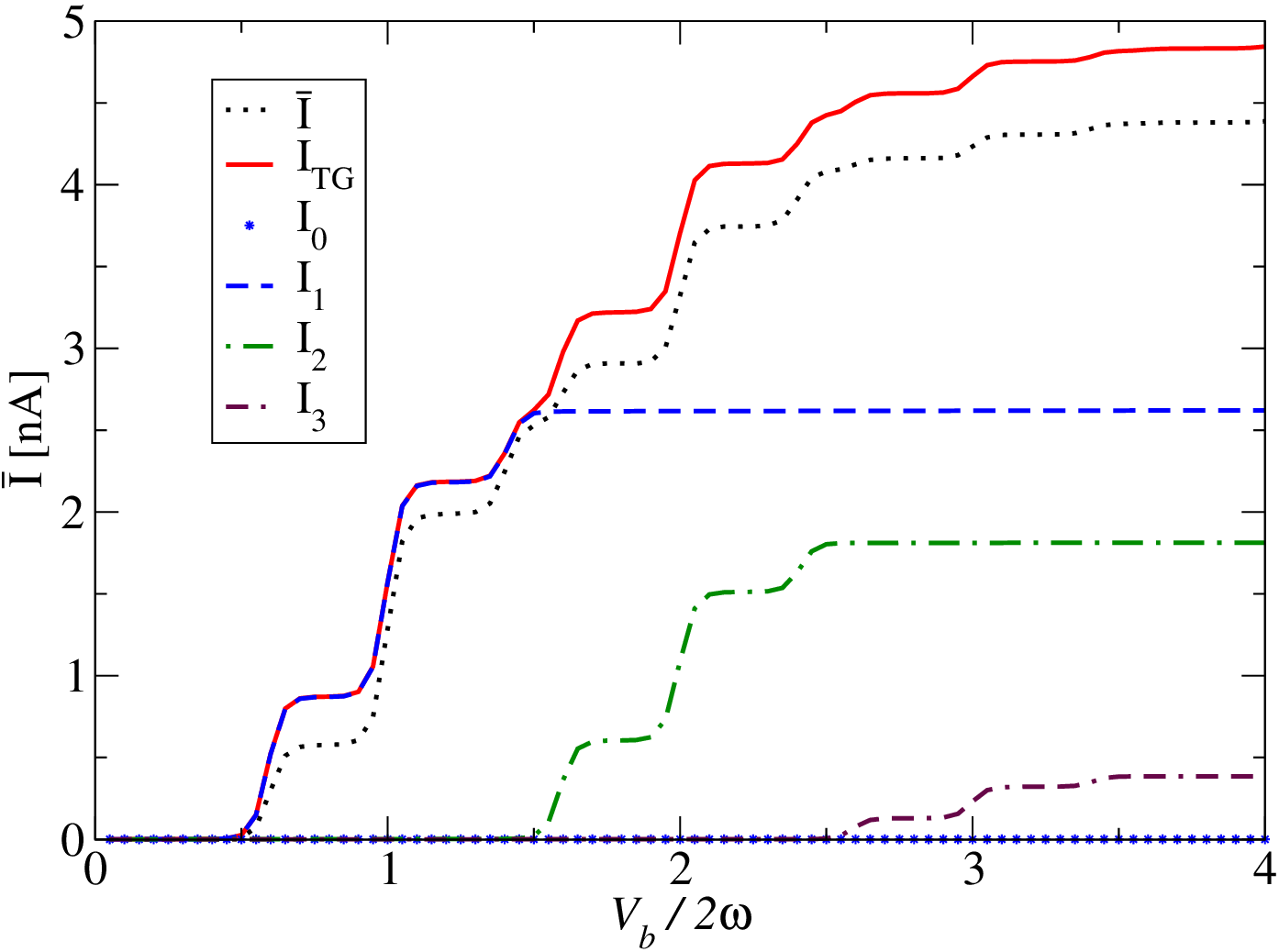}}
\caption{Same as Fig.~1 but with electron interaction $U$=$0.4\omega$.
Shoulders in the current are at $V_b/ (2 \omega)$=0.6, 1.0, 1.4, 1.6, 2.0, etc.
}
\label{f.3}
\end{figure}

\begin{figure}
\centerline{
\includegraphics[width=8.5cm,clip]{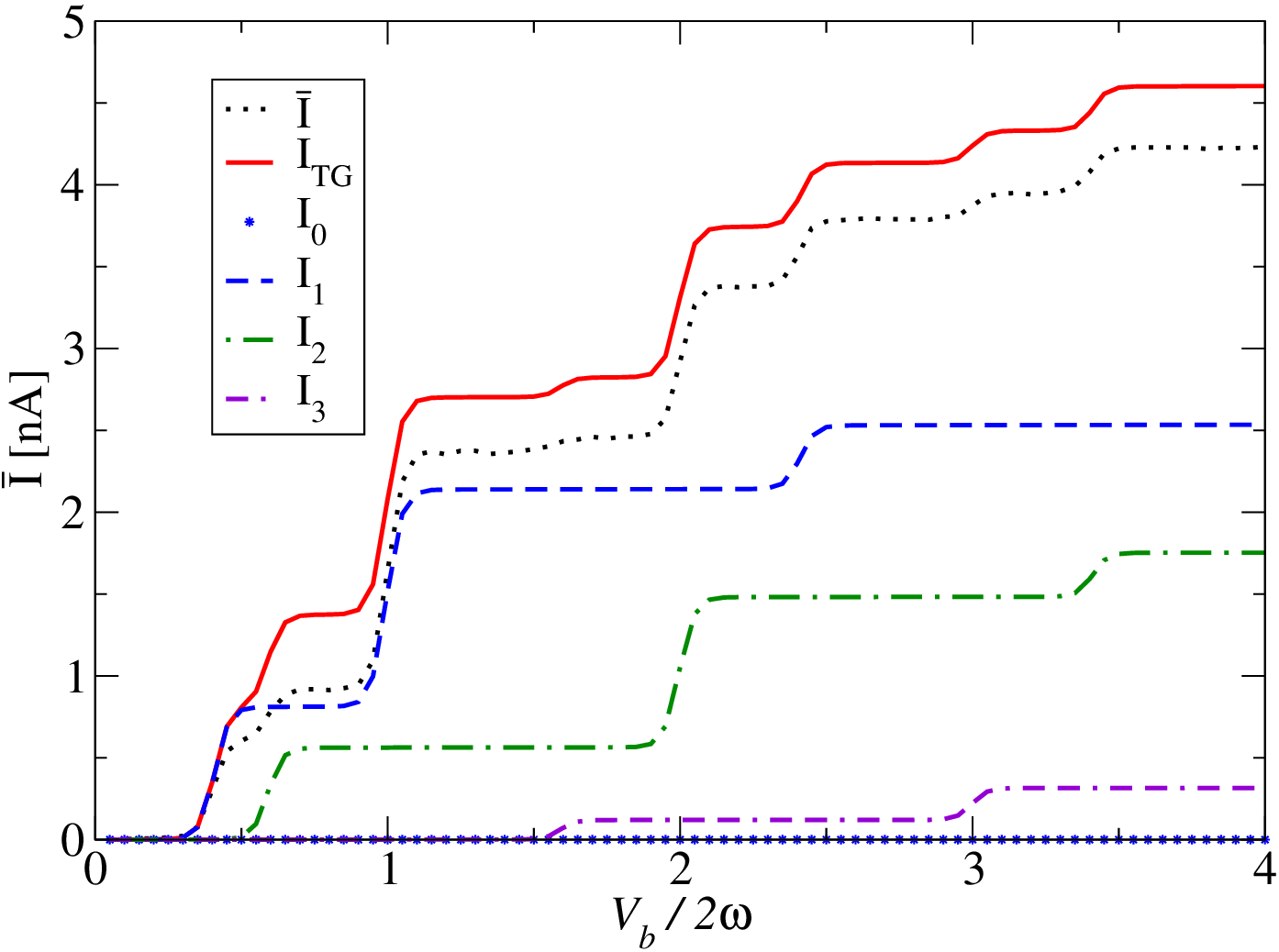}}
\caption{Same as Fig.~1 but with electron interaction $U$=$1.4\omega$.
  Shoulders in the current are at $V_b/ (2 \omega)$=0.4, 0.6, 1.0, 1.4,
  1.6, 2.0, etc.  }
\label{f.31}
\end{figure}

At first sight it might be astonishing that CDT also works for one site
coupled to reservoirs since the infinitely large reservoirs could be
expected to destroy any coherent effect. Let us go back to the two-level
system \cite{gros91,gros92} in which it does not matter whether both or
just one level is periodically driven to be able to observe CDT. It is also
not important if the static level is replaced by a continuum of states,
since the driving only modifies the coupling. As can be seen from the
present results the situation is even not changed if the continuum of
states is replaced by a reservoir or if a second one is added. In all these
situations the coupling between the sites can be suppressed if the correct
amplitude of the external field is applied.

%To see this more clearly we investigated a related three-site system
%without coupling to any reservoirs.  Without external field all sites have
%the same energy and the field only moves the energy of the middle site up
%and down periodically. The population dynamics is shown in Fig. \ref{f.2}.
%When the CDT condition is fulfilled, tunneling is suppressed as in the
%two-site case \cite{gros91,gros92}. The molecular junction system is
%similar to this test system in that the energies of the leads are not
%changed and the field just influences the molecule in the middle. The right
%and left sites can be seen as very simplistic versions of the reservoirs in
%the molecular junction.

%\vspace{3ex} 

\section{Monochromatic laser,   \bm{$0<U<\omega$}}
%\section{Case $0<U<\omega$}
 Let us first concentrate on this case of small
Coulomb interaction.  When the bias
voltage $V_b$ is small, the current is established through the channel with
energy $\varepsilon_0$.  As soon as $V_b/2$ approaches $\omega$, an extra
channel with energy $\varepsilon_0+U$ is opened by the electron interaction
and a step in the current-voltage characteristics appears.  Under the
influence of the external field, the PAT induces additional current
channels, i.e.\ states with energies $\varepsilon_0\pm{}n\omega$ and
$\varepsilon_0+U\pm{}n\omega$.  As mentioned above, the bias is increased
symmetrically with respect to $\varepsilon_0$, i.e.\
$E_{F,l}=\varepsilon_0+V_b/2$ and $E_{F,r}=\varepsilon_0-V_b/2$. So a step
in the $I$-$V_b$ curve appears whenever an additional conduction channel
can participate in the tunneling.  Therefore one finds steps at $V_b/2$
values $n\omega-U$, $n\omega$ and $n\omega+U$.  In Fig.~\ref{f.3} with
$U$=0.4 $\omega$ one observes respective shoulders at $V_b/ (2
\omega)$=0.6, 1.0, 1.4, 1.6, 2.0, etc. The CDT phenomenon occurs again for
small bias voltages and disappears at $V_b/2=0.6 \omega$. The partial
current $I_0$ has, in principle, a shoulder at $V_b/2=0.4\omega$ but it is
not visible in Fig.~\ref{f.3} because $I_0=0$ due to CDT.

%\section{Case $U>\omega$}
\section{Monochromatic laser,   \bm{$U>\omega$}}
In this case the current-voltage curve behaves similar to that for $0<U<
\omega$.  Current steps still appear at $V_b/2$ values $n\omega$ and
$n\omega+U$ when the energy levels match the higher (left) Fermi level.
The energy levels $\varepsilon_0+U-n \omega$ of some conduction channels
can be larger than $\varepsilon_0$ and might coincide with the higher Fermi
level, but the states with $\varepsilon_0-n \omega$ have again to be
compared to the lower (right) Fermi energy. As shown in Fig.~\ref{f.31} for
$U$=1.4 $\omega$ one observes the first shoulder at $V_b/ (2 \omega)$=0.4
and this step relates to the channel energy $\varepsilon_0+U-\omega$.
Further shoulders can be observed at $V_b/ (2 \omega)$=0.6, 1.0, 1.4, 1.6,
2.0, etc.  All energetic step positions except the one at $V_b/ (2
\omega)$=0.4 are the same as in Fig.~\ref{f.3} due to the fact that the
electron interaction strength $U$ differs by the driving frequency $\omega$
between these two figures. The heights of the individual steps in the
current differ considerably as does the physical reasoning for the steps.
This is best seen if one looks at the partial currents $I_1$, $I_2$, and
$I_3$ which constitute the Tien-Gordon current.  For the large electron
interaction case the current $I_2$, for example, sets in much earlier.  The
Coulombic staircases within the partial currents $I_n$ have of course a
width depending on the size of $U$.  Similar studies can be performed for
an attractive interaction $U<0$.

\begin{figure}
\centerline{
\includegraphics[width=8.5cm,clip]{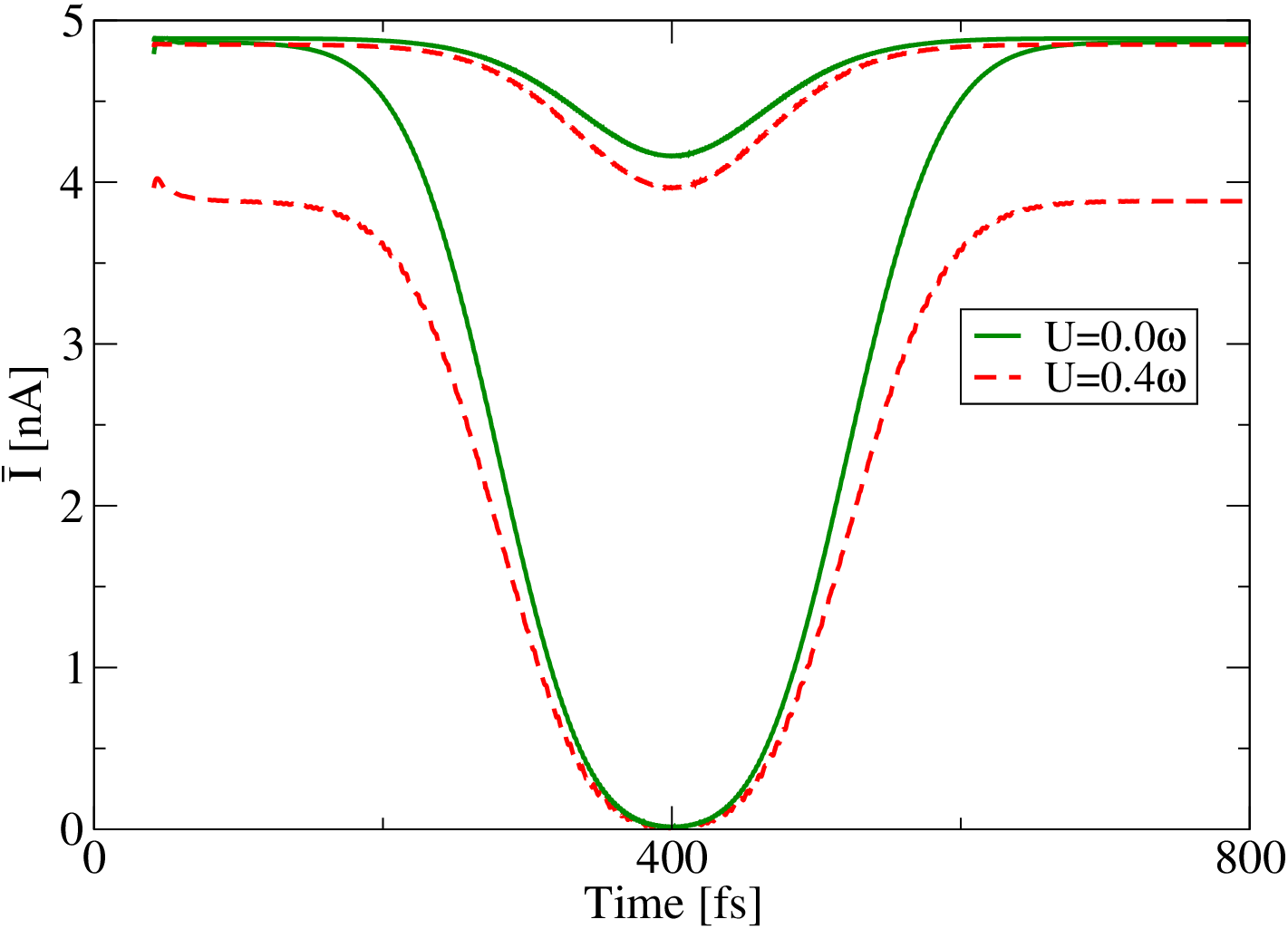}}
\caption{Current induced by laser with Gaussian shape of the envelope with
  amplitude $E_0=2.405 \omega \exp\left(-(t-t_0)^2/(2\sigma^2)\right)$ with
  $t_0$=400 fs and $\sigma$=80 fs.  Average current $\bar{I}$ vs time for
  $k_B T$=$0.025\omega$. The two top lines are for a bias
  $V_b/2$=$2.4\omega$. The two lower lines are for a small bias
  $V_b/2$=$0.4\omega$.  The solid lines represent $U$=0, the dashed lines
  $U$=$0.4\omega$.}
\label{f.4}
\end{figure}

\section{Laser pulse} Above we investigated laser fields with constant
amplitude $E_0=2.405\omega$, i.e.\ equal to a zero of the Bessel function
$J_0$. One can also use other external fields with time-dependent envelope
function, e.g.\ with a Gaussian envelope $E_0(t)=2.405\omega
\exp\left(-(t-t_0)^2/(2\sigma^2)\right)$ \cite{klei06b}. For the following
example we set $t_0$=400 fs and $\sigma$=80 fs. Since the amplitude of this
laser is time-dependent, it induces many more inelastic channel states.
During the pulse the laser amplitude changes $E_0$ from 0 to $2.405\omega$
following the Gaussian shape.  At each moment in time there is one
resonance state $\varepsilon_0$ and many PAT-induced energy states
$\varepsilon_0\pm{}n \omega$ with probability $J_n^2(E_0/\omega)$. For a small
bias voltage only the partial current $I_0$ is non-zero and all other
partial currents vanish. So in this case the fulfillment of the CDT
condition leads to a complete suppression of the current for $t=t_0$. For
larger bias, the CDT condition just suppresses the partial current $I_0$
but not the total current. In Fig.~\ref{f.4}, when the bias is small, CDT
can be seen clearly with and without electron interaction.  When the bias
is big, CDT is only visible as a small dip in the current.  For the case
with electron interaction $U=0.4\omega$, the energy level $\varepsilon_0+U$
just lines up with the Fermi level for the bias $0.8\omega$, and the
current tunneling through this channel is smaller than the resonance
current. This is the reason why the current is somewhat smaller than
without electron interaction.

As mentioned above, the average current $\bar{I}$ was obtained by averaging
$I(t)$ over five periods of the highly oscillating carrier field. For the
laser pulse with time-dependent envelope the current $\bar{I}$ is of course
depending on the number $n_{av}$ of periods used in the averaging
procedure. In these cases any averaging procedure will not only average
over the highly-oscillating carrier field but also smooth the envelope
function. This effect will become more pronounced for large $n_{av}$.  For
a Gaussian shape of the laser pulse the averaged current will still be
close to a Gaussian form although slightly shifted and unsymmetric. For the
example in Fig.~\ref{f.4} there are only minor changes when changing
$n_{av}$ between 2 and 10. The averaging process does not depend much on
$n_{av}$ for monochromatic laser pulses.

\section{Conclusions} To conclude, we have shown that the phenomenon of CDT
already appears in molecular junctions which can be modeled by one site
coupled to two contacts and not only in systems in which the tunneling is
quenched within the molecule itself. Since the total current consists of
several partial currents additional channels might be opened by the
external field. A complete suppression of the current occurs only for small
bias voltages when only the channel for $I_0$ is open and if the amplitude
of the external field fulfills the CDT condition. In the present letter we
concentrated on the case of a high carrier frequency but as in the case of
the two-level system similar effects should be possible for smaller
frequencies \cite{cref04b}. The density matrix formalism used here
restricts the present results to weak coupling between the leads and the
molecular junction. In the case of CDT between two sites the phenomenon
does not depend on the size of the coupling. Therefore we believe that the
effect of CDT between a single site coupled to one or more leads will also
survive for stronger coupling between site and leads.  Although the present
model calculations certainly can only be a very rough description of real
systems, the hope is that the basic physics survives in more realistic and
complex systems.  Then the effect of CDT with short laser pulses might
allow for the construction of fast opto-electronic switches if one finds
materials with long enough coherence times.

The authors would like to thank the DFG for financial support within SPP
1243.

%\bibliographystyle{epl}
%\bibliography{ukleine}

\begin{thebibliography}{10}

\bibitem{hang02b}
\Name{H\"anggi~P., Ratner~M., \and Yaliraki~S. (Eds.)} 
\REVIEW {Chem. Phys.}{281}{2002}{111}
special issue on transport in molecular wires.

\bibitem{nitz03a}
\Name{Nitzan~A. \and Ratner~M.~A.}
\REVIEW {Science}{300}{2003}{1384}.

\bibitem{joac04}
\Name{Joachim~C. \and Ratner~M.~A.}
\REVIEW {Nanotechnology}{15}{2004}{1065}.

\bibitem{plat04a}
\Name{Platero~G. \and Aguado~R.}
\REVIEW {Phys. Rep.}{395}{2004}{1}.

\bibitem{daye62}
\Name{Dayem~A. \and Martin~R.}
\REVIEW {Phys. Rev. Lett.}{8}{1962}{246}.

\bibitem{tien63}
\Name{Tien~P. \and Gordon~J.}
\REVIEW {Phys. Rev.}{129}{1963}{647}.

\bibitem{gros91}
\Name{Grossmann~F., Dittrich~T., Jung~P., \and H\"anggi~P.}
\REVIEW {Phys. Rev. Lett.}{67}{1991}{516}.

\bibitem{gros91b}
\Name{Grossmann~F., Jung~P., Dittrich~T., \and H\"anggi~P.}
\REVIEW {Z. Phys. B}{84}{1991}{315}.

\bibitem{gros92}
\Name{Grossmann~F. \and H\"anggi~P.}
\REVIEW {Europhys. Lett.}{18}{1992}{571}.

\bibitem{lehm03a}
\Name{Lehmann~J., Camalet~S., Kohler~S., \and H\"anggi~P.}
\REVIEW {Chem. Phys. Lett.}{368}{2003}{282}.

\bibitem{cama03}
\Name{Camalet~S., Lehmann~J., Kohler~S., \and H\"anggi~P.}
\REVIEW {Phys. Rev. Lett.}{90}{2003}{210602}.

\bibitem{cama04}
\Name{Camalet~S., Kohler~S., \and H\"anggi~P.}
\REVIEW {Phys. Rev. B}{70}{2004}{155326}.

\bibitem{kohl04b}
\Name{Kohler~S., Camalet~S., Strass~M., Lehmann~J., Ingold~G.~L., \and
  H\"anggi~P.}
\REVIEW {Chem. Phys.}{296}{2004}{243}.

\bibitem{kohl04a}
\Name{Kohler~S., Lehmann~J., \and H\"anggi~P.}
\REVIEW {Phys. Rep.}{406}{2005}{379}.

\bibitem{wela05a}
\Name{Welack~S., Schreiber~M., \and Kleinekath\"ofer~U.}
\REVIEW {J. Chem. Phys.}{124}{2006}{044712}.

\bibitem{bran00}
\Name{Brandes~T. \and Renzoni~F.}
\REVIEW {Phys. Rev. Lett.}{85}{2000}{4148}.

\bibitem{cref04}
\Name{Creffield~C.~E. \and Platero~G.}
\REVIEW {Phys. Rev. B}{69}{2004}{165312}.

\bibitem{cota05}
\Name{Cota~E., Aguado~R., \and Platero~G.}
\REVIEW {Phys. Rev. Lett.}{94}{2005}{107202}.

\bibitem{galp05}
\Name{Galperin~M. \and Nitzan~A.}
\REVIEW {Phys. Rev. Lett.}{95}{2005}{206802}.

\bibitem{klei06b}
\Name{Kleinekath\"ofer~U., Li~G.-Q., Welack~S., \and Schreiber~M.}
\REVIEW {Europhys. Lett.}{75}{2006}{139}.

\bibitem{klei06c}
\Name{Kleinekath\"ofer~U., Li~G.-Q., Welack~S., \and Schreiber~M.}
\REVIEW {phys.\ stat.\ sol.~(b)}{243}{2006}{3775}.

\bibitem{thie03}
\Name{Thielmann~A., Hettler~M.~H., K\"onig~J., \and Sch\"on~G.}
\REVIEW {Phys. Rev. B}{68}{2003}{115105}.

\bibitem{brud94}
\Name{Bruder~C. \and Schoeller~H.}
\REVIEW {Phys. Rev. Lett.}{72}{1994}{1076}.

\bibitem{staf96}
\Name{Stafford~C.~A. \and Wingreen~N.}
\REVIEW {Phys. Rev. Lett.}{76}{1996}{1916}.

\bibitem{oost96}
\Name{Oosterkamp~T.~H., Kouwenhoven~L.~P., Koolen~A. E.~A., van~der
  Vaart~N.~C., \and Harmans~C. J. P.~M.}
\REVIEW {Semicond. Sci. Technol.}{11}{1996}{1512}.

\bibitem{oost97}
\Name{Oosterkamp~T.~H., Kouwenhoven~L.~P., Koolen~A. E.~A., van~der
  Vaart~N.~C., \and Harmans~C. J. P.~M.}
\REVIEW {Phys. Rev. Lett.}{78}{1997}{1536}.

\bibitem{oost98}
\Name{Oosterkamp~T.~H.}
\REVIEW {Nature}{395}{1998}{873}.

\bibitem{sun98}
\Name{Sun~Q.~F., Wang~J., \and Lin~T.~H.}
\REVIEW {Phys. Rev. B}{58}{1998}{13007}.

\bibitem{qin01}
\Name{Qin~H., Simmel~F., Blick~R.~H., Kotthaus~J.~P., Wegscheider~W., \and
  Bichler~M.}
\REVIEW {Rep. Prog. Phys.}{63}{2000}{669}.

\bibitem{bran04}
\Name{Brandes~T., Aguado~R., \and Platero~G.}
\REVIEW {Phys. Rev. B}{69}{2004}{205326}.

\bibitem{bran05}
\Name{Brandes~T.}
\REVIEW {Phys. Rep.}{408}{2005}{315}.

\bibitem{bran06}
\Name{Brandes~T.}
\REVIEW {phys. stat. sol. (b)}{243}{2006}{2293}.

\bibitem{cref04b}
\Name{Creffield~C.~E.}
\REVIEW {Europhys. Lett.}{66}{2004}{631}.

\end{thebibliography}

\end{document}